\documentstyle[preprint,aps,epsfig]{revtex}
\begin{document}
\draft
\title{Wetting at Non-Planar Substrates: Unbending \& Unbinding}

\author{C.\ Rasc\'{o}n, A.O.\ Parry, A.\ Sartori}
\address{Mathematics Department, Imperial College\\
180 Queen's Gate, London SW7 2BZ, United Kingdom}
\date{\today}
\maketitle

\begin{abstract}
We consider fluid wetting on a corrugated substrate using
effective interfacial Hamiltonian theory and show that breaking the
translational invariance along
the wall can induce an {\it unbending } phase transition in
addition to unbinding. Both first order and second order
unbending transitions can occur at and out of coexistence.
Results for systems with short-ranged and long-ranged
forces establish that the unbending critical point is
characterised by hyperuniversal scaling behaviour. We
show that, at bulk coexistence,
the adsorption at the unbending critical point
is a universal multiple of the adsorption for the correspondent
planar system.
\end{abstract}

Recently, the subject of fluid adsorption and wetting on
structured (non-planar) and heterogeneous substrates has began to
receive considerable attention \cite{Rough}. This work is not only
a natural extension of studies of wetting on idealised planar
surfaces \cite{Dietrich} but it is also of more fundamental interest since the
broken translational invariance along the wall necessarily leads to
competition between surface tension and direct molecular effects.
Thus, we may anticipate that new interesting phenomena
(phase transitions, scaling, universality) will emerge which
do not occur for planar systems. In this letter, we report
results of extensive numerical calculations, supported by approximate
non-perturbative analysis and scaling theory, of wetting on a
periodic (corrugated) substrate. These reveal that novel
first and second order transitions can take place,
directly related to the inhomogeneity along the wall.
For long-ranged forces,
the phase transition, referred as unbending, only occurs
for sufficiently large wall corrugations (beyond the range
of previously employed perturbative methods \cite{Perturbation}),
dependent on the wave vector of the corrugation.
In contrast, for short-ranged forces, the critical threshold
is wave vector independent and rather weak.
There are three aspects of our work that we emphasize
in particular. Firstly, the unbending transition precedes a
wetting (unbinding) transition occurring at a higher temperature
(and at bulk two-phase coexistence).
For second order unbinding transitions, on which we concentrate,
the location of the wetting transition is unaffected by wall
corrugation.
Secondly, the location of the unbending line and critical point,
as well as the interface structure, only depend on the amplitude and
period of the wall corrugation function through hyperuniversal
scaling variables analogous to that encountered in the theory
of finite-size effects at bulk critical points \cite{Cardy}.
As a consequence,
the unbending critical point is associated with non-trivial
universal amplitude ratios which relate the adsorptions in the
non-planar and correspondent planar system. Finally, unbending
is directly related to nonlinear bifurcation
phenomena occurring in dynamical systems, a subject whose
mathematical aspects continue to attract attention \cite{Equation}.

To begin, we describe the results of a specific mean-field (MF)
model of unbending and unbinding which also serves to illustrate
important scaling properties which we shall later put in a more
general context. For simplicity, we assume that the wall has 
a corrugated sinusoidal shape $\psi(x)\!=\!a\cos({\sl q}x)$, which
breaks the translational invariance in one direction only.
Following the work of earlier authors \cite{Rough,Perturbation},
we take
as our starting
point the (reduced) standard effective interfacial model
\begin{equation}
\label{one}
H[\ell]={1\over L}\int_{L}\!dx
\;\left[\,{\Sigma\over 2}\left({{\partial\ell}\over{\partial x}}\right)^2
+W(\ell-\psi)\,\right]
\end{equation}
restricted to the space of periodic solutions which is sufficient
for our description of equilibrium phenomena. Here, $\Sigma$ is the
surface stiffness, $W$ is the binding potential and $\ell(x)$ is the
collective coordinate measuring the height of the interface 
relative to the mean position of the wall whose period $L$
satisfies ${\sl q}=2\pi/L$. We
also restrict ourselves to a MF description in which the
equilibrium profiles $\ell_{\nu}$ are obtained by minimising Eq.\
(\ref{one}). The importance of fluctuation effects will be discussed
later in the context of scaling theory \cite{Bau}. We start by considering
systems with short-ranged forces at bulk two-phase coexistence
and write \cite{Dietrich}
\begin{equation}
\label{two}
W(\ell)=-\Delta T\;e^{-\ell}+\beta\;e^{-2\ell}\; ;
\end{equation}
so that both the film thickness $\ell$ and corrugation amplitude
$a$ are measured
in units of the bulk correlation length.
With this potential (and positive $\beta$)
the planar system undergoes a second-order unbinding transition
at $\Delta T\!\equiv\! T_w\!-\!T\!=\!0$ such that the MF interface thickness
and the transverse correlation length
diverge at that critical point as
$\ell_{\pi}\sim-\log(\Delta T)$ and $\xi_\parallel\sim
\Delta T^{\;-1}$, corresponding to standard wetting critical
exponents $\beta_{\sf S}\!=\!0\,(\log)$ and $\nu_\parallel\!=\!1$
respectively \cite{Dietrich}.

For $a\neq 0$, the MF
configuration(s) are the solutions of the Euler-Lagrange equation
\begin{equation}
\Sigma\;\ell_{\nu}''(x)=W'(\ell_{\nu}-\psi),
\label{three}
\end{equation}
solved subject to periodic boundary conditions and where
the prime denotes differentiation  w.r.t.\ the argument.
This deceptively simple looking nonlinear equation can show
multiple solutions and bifurcations corresponding to different possible
phases for the equilibrium interface configuration. Whilst a full
analytic solution is not possible, it is straightforward to
show that the solutions exhibit an important scaling property which allows
us to collapse results obtained for different periods $L\!\!=\!\!2\pi/{\sl q}$
onto a universal surface phase diagram. To see this, we introduce
the new variables $\eta\!\equiv\!\ell-\psi-\ell_{\pi}$ and
$t\!\equiv\!{\sl q}\,x$ so that (\ref{three}) becomes
\begin{equation}
\ddot{\eta}=\Delta \widetilde{T}^{\;\,2}\,(\,e^{-\eta}-e^{-2\eta})
+a\,\cos t,
\label{five}
\end{equation}
which is the equation of a forced inverted nonlinear oscillator.
Here the overdot corresponds to differentiation w.r.t.\ $t$ whilst
the temperature, stiffness and substrate periodicity are
combined in the rescaled temperature variable $\Delta\widetilde{T}\!\equiv\!
\Delta T/{\sl q}\sqrt{2\beta\Sigma}$. Consequently,
any new phase transition induced by the corrugation amplitude $a$
is not affected by the value of the wall periodicity $\sl q$ which
only acts to rescale the temperature deviation from $T_w$. In Fig.\ 1,
we show plots of the mean interface thickness $\ell_0$,
defined as the average $<\!\ell(x)\!>_{x}$, as a function of
$\Delta\widetilde{T}$ for various $a$,
obtained by numerically minimising Eq.\ (\ref{one}).
It can be seen that, whilst
the location of the unbinding transition is unaffected by the wall
corrugation, a new phase transition occurs for corrugation amplitudes
$a\!>\!a_{\sf c}\!\approx\! 2.914$ and $\Delta\widetilde T\!>\!
\Delta\widetilde T_{\sf c}\!\approx\! 2.12$.
The surface phase diagram is shown in Fig.\ 2
and exhibits the termination of the first-order phase
boundary at an unbending critical point as well as
representative shapes of the coexisting interfacial phases
at the transition. Again, we emphasize the universal value of the
critical corrugation amplitudes $a_{\sf c}$ (which is independent of $\sl q$)
whilst the temperature shift from $T_{w}$ satisfies
$\Delta T_{\sf c}({\sl q})\propto {\sl q}$.

Before we discuss further scaling properties that emerge from the exact
minimization of (\ref{one}), we describe an approximate treatment of
the model which recovers the unbending transition and yields
relatively good values for the critical point.
To this end, we suppose that the interface
configuration, and consequently the free-energy, can be parametrized
by two variables by restricting ourselves to profiles of the form
$\ell(x)\approx\ell_0+(1-\epsilon)\psi(x)$. Thus, $\ell_0$ is the average
interface displacement whilst $\epsilon$ measures the extent of
interfacial corrugation. The bounding value $\epsilon\!=\!1$
corresponds to a completely flat configuration whereas
$\epsilon\!=\!0$
refers to a configuration with identical corrugation to the
wall. Substituting this parametrized profile shape into the
Hamiltonian, Eq.\ (\ref{one}), and minimising w.r.t.\ $\ell_0$, we are led
to the following approximate expression for the dependence of the
free-energy $F$ on the interface corrugation parameter $\epsilon$,
\begin{equation}
{2\over{\Sigma{\sl q}^2}}\,F(\epsilon)={{a^2}\over 2}(1-\epsilon)^2
\,-\,\Delta\widetilde T^{\;\,2}\,{{I_{0}^{2}(\epsilon a)}\over
{I_{0}(2\epsilon a)}}
\label{six}
\end{equation}
where $I_{0}$ denotes the modified Bessel function of zero-order.
The two terms on the r.h.s.\ represent the competition between the surface
tension and binding potential effects which are each minimized separately
by $\epsilon\!=\!1$ (flat interface) and $\epsilon\!\!=\!\!0$
(corrugated interface) respectively.
Plots of $F(\epsilon)$ for various $a$ moving along
the unbending line are shown in Fig.\ 3 and illustrate
the possibility of phase coexistence between bent and rather
flat states for sufficiently large $a$. The locus of the unbending
transition in the surface phase diagram obtained in this approximate
manner is shown as the dashed line in Fig.\ 2 and agrees
reasonably well with the exact numerical result. Note that the solutions
will only depend on $\Delta\widetilde{T}$ and $a$, as in the exact
solution. This method also has
a distinct advantage over previously adopted perturbative treatments
\cite{Perturbation}
(involving an expansion about the planar system) which, whilst not
without merit, cannot handle the occurrence of distinct branches
({\it i.e.\ }a bifurcation) in the free-energy \cite{Oops}.
We also note that the location of the unbending critical point
within this approximate non-perturbative method can be determined
with an elegant graphical construction \cite{RPS}.

We consider now the same phenomena for systems with
long-ranged (dispersion) forces. For this case, we use the
binding potential \cite{Dietrich}
\begin{equation}
\label{seven}
W(\ell)=-{{\Delta T}\over{\ell^2}}+{\beta\over{\ell^3}}\;
\end{equation}
which again describes a continuous unbinding transition in the planar system
as $T\rightarrow T_w$ \cite{Dietrich}.
For this system, the film thickness and transverse correlation length
diverge as $\ell_{\pi}\sim\Delta T^{\;-1}$ and $\xi_\parallel\sim
\Delta T^{\;-5/2}$, corresponding to critical
exponents $\beta_{\sf S}\!=\!1$ and $\nu_\parallel\!=\!5/2$
respectively \cite{Dietrich}.
Turning to the
non-planar geometry, we make the judicial change of variables
$\eta\!\equiv\!(\ell-\psi)/\ell_{\pi}$ and
$t\!\equiv\!{\sl q}\,x$
which again reduces the Euler-Lagrange equation (\ref{three}) to
that of a forced inverted nonlinear oscillator:
\begin{equation}
\ddot{\eta}=\Delta\widetilde T^{\;\,2}\,(\,{1\over{{\eta}^3}}
-{1\over{{\eta}^4}})
+\tilde{a}\,\cos t\, .
\label{nine}
\end{equation}
Once more, the two scaling variables
$\Delta\widetilde{T}\!\equiv\!2\,\Delta T/\Sigma{\sl q}^{2}\ell_{\pi}^{4}$
and $\widetilde a\!\equiv\!a/\ell_{\pi}$ determine the multiplicity
of solutions and hence the surface phase diagram.

Plots of the mean interface position $\ell_{0}$
{\it vs.\ }$\Delta\widetilde{T}$ for
different $a$ obtained from the numerical minimization of
Eq.\ (\ref{one}) are, in essence, the same as that shown in Fig.\ 1
for short-ranged forces and, therefore, are not presented here.
The numerical values for the scaled variables at the
unbending critical point are $\widetilde{a}_{\sf c}\!\approx\!2.061$ and
$\Delta\widetilde{T}\!\approx\!8.66$ which imply a wave-vector dependence
$a_{\sf c}({\sl q})\propto {\sl q}^{-2/5}$ and
$\Delta T_{\sf c}({\sl q})\propto {\sl q}^{2/5}$ for the
critical corrugation amplitude and temperature shift, respectively.

The MF results described above suggest that the location of the
unbending critical point can be understood using scaling theory.
To this end, we suppose that, in the planar system, the excess
free-energy per unit area contains a singular contribution
$F_{\pi}^{\hbox{sing}}\sim\Delta T^{\,2-\alpha_{\sf S}}$
(with $\alpha_{\sf S}\!=\!0$ and $-1$ for the model potential
(\ref{two}) and (\ref{seven}), respectively \cite{Dietrich}).
In the non-planar system, we conjecture that the corresponding
quantity is described by the scaling function
\begin{equation}
\Delta F_{\nu}^{\hbox{sing}}=\Delta T^{\;2-\alpha_{\sf S}}\;\;
W\left(a\,\Delta T^{\;\beta_{\sf S}},{\sl q}\,
\Delta T^{\;-\nu_\parallel}\right)
\label{ten}
\end{equation}
where $W(x,y)$ is the scaling function whose variables correspond
to the hyperuniversal combination of lengthscales
$a/\ell_{\pi}$ and ${\sl q}\,\xi_{\parallel}$ \cite{Note}.
Since the singularity in the free-energy at the unbending critical
point occurs for $\Delta T\!\neq\!0$, we are immediately led to
the prediction for the critical corrugation amplitude and temperature
\begin{equation}
a_{\sf c}({\sl q})\propto {\sl q}^{-{\beta_{\sf S}\over
\nu_{\parallel}}}\,;\hspace{.6cm}
\Delta T_{\sf c}({\sl q})\propto {\sl q}^{1\over{\nu_{\parallel}}}
\label{eleven}
\end{equation}
consistent with our explicit results,
provided that for short-ranged forces we interpret
$\beta_{\sf S}/\nu_{\parallel}$ as zero and not logarithmic.
We believe that the existence of a finite critical threshold
even in the $q\!\rightarrow\!0$ limit is a surprising
finding of our work.
These scaling ideas can be extended to the interface structure
at the unbending critical point where the hyperuniversal nature
of the scaling variables $x$ and $y$ play an important role.
Here, we concentrate on systems with long-ranged forces for which
$\beta_{\sf S}\!\neq\!0$ where the definition of universal critical
amplitudes is more straightforward. We suppose that, in the vicinity
of the unbending critical point, the mean interface thickness in the
non-planar system is described by the scaling law
\begin{equation}
\ell_{0}=\;\ell_{\pi}\;
\Lambda\left({a\over{\ell_{\pi}}},{\sl q}\,\xi_{\parallel}\right)
\label{twelve}
\end{equation}
where $\Lambda(x,y)$ is a universal scaling function.
As a consequence, precisely at the unbending critical point, the mean
film thickness $\ell_{0}^{\,\sf c}$ is a universal multiple of the
corresponding planar adsorption (at the same temperature). Thus,
we define the universal critical amplitude ratio
\begin{equation}
R\equiv{{\ell_{0}^{\,\sf c}}\over{\ell_{\pi}}}
\hspace{.5cm}\hbox{at}\hspace{.5cm}a\!=\!a_{\sf c}({\sl q}),\,
\Delta T\!=\!\Delta T_{\sf c}({\sl q})
\label{thirteen}
\end{equation}
which we have numerically determined as $R\!\approx\!1.321$ (independent
of {\sl q}) calculated using our MF theory with the binding potential
(\ref{seven}). Note that the definition of $R$ is equivalent to
the ratio of adsorptions in the non-planar and planar systems.
Other universal critical amplitudes can also be defined. For example,
at the unbinding critical point, the shift in the mean interface
height relative to the planar system satisfies
\begin{equation}
R'\equiv{{\ell_{0}^{\,\sf c}-\ell_{\pi}(\Delta T_{\sf c})}
\over{a_{\sf c}({\sl q})}}
\label{fourteen}
\end{equation}
with $R'$ also independent of $\sl q$. The advantage of this
definition is that it is also appropriate for systems in which
$\beta_{\sf S}\!=\!0\;(\log)$. We have numerically determined that
$R'=0.640$ and $0.156$ for the potentials
(\ref{two}) and (\ref{seven}) respectively.

To finish our article, we make two pertinent remarks. Firstly,
we have established that for $a>a_{\sf c}$ the first order
unbending transition also occurs out of the two-phase coexistence
for sufficiently small bulk ordering field $\bar{h}$. The
result of our numerical calculations for short-ranged forces
including an additional $\bar{h}\ell$ term in the binding potential
are shown in Fig.\ 4. The existence of an unbending line
extending out of bulk two-phase coexistence is analogous
to prewetting at (planar) first-order phase transitions.
Secondly, we have established that unbending also occurs for
first order wetting transitions in non-planar systems although the
scaling behaviour is less obvious. A section of the surface phase
diagram in the $(T,\bar{h})$ plane thus shows both prewetting
and unbending lines.
While this first appears similar to prefilling on a wedge,
there are profound and subtle differences between unbending
and prefilling relating to the order of these transitions
and their relation with wetting.
In summary, we have shown that for
non-planar systems an additional interfacial phase transition
is associated with unbinding.
The critical point of the unbending transition exhibits novel
scaling and observable universal critical properties. Further
work should concentrate on more general wall shapes, calculations
with more microscopic models and also aim to
establish whether the values of the universal critical
amplitudes presented here are substantially affected by
including fluctuation effects beyond mean-field level.
At present, simulation studies seem best equipped to answer
this latter question although renormalization group
analysis may be possible.

C.R. is on leave from the Departamento de F\'{\i}sica Te\'{o}rica
de la Materia Condensada, Universidad Aut\'{o}noma de Madrid,
and acknowledges economical support from {\it La Caixa} and
The British Council.

\begin{figure}[p]
\label{first}
\vspace*{3cm}
\epsfig{file=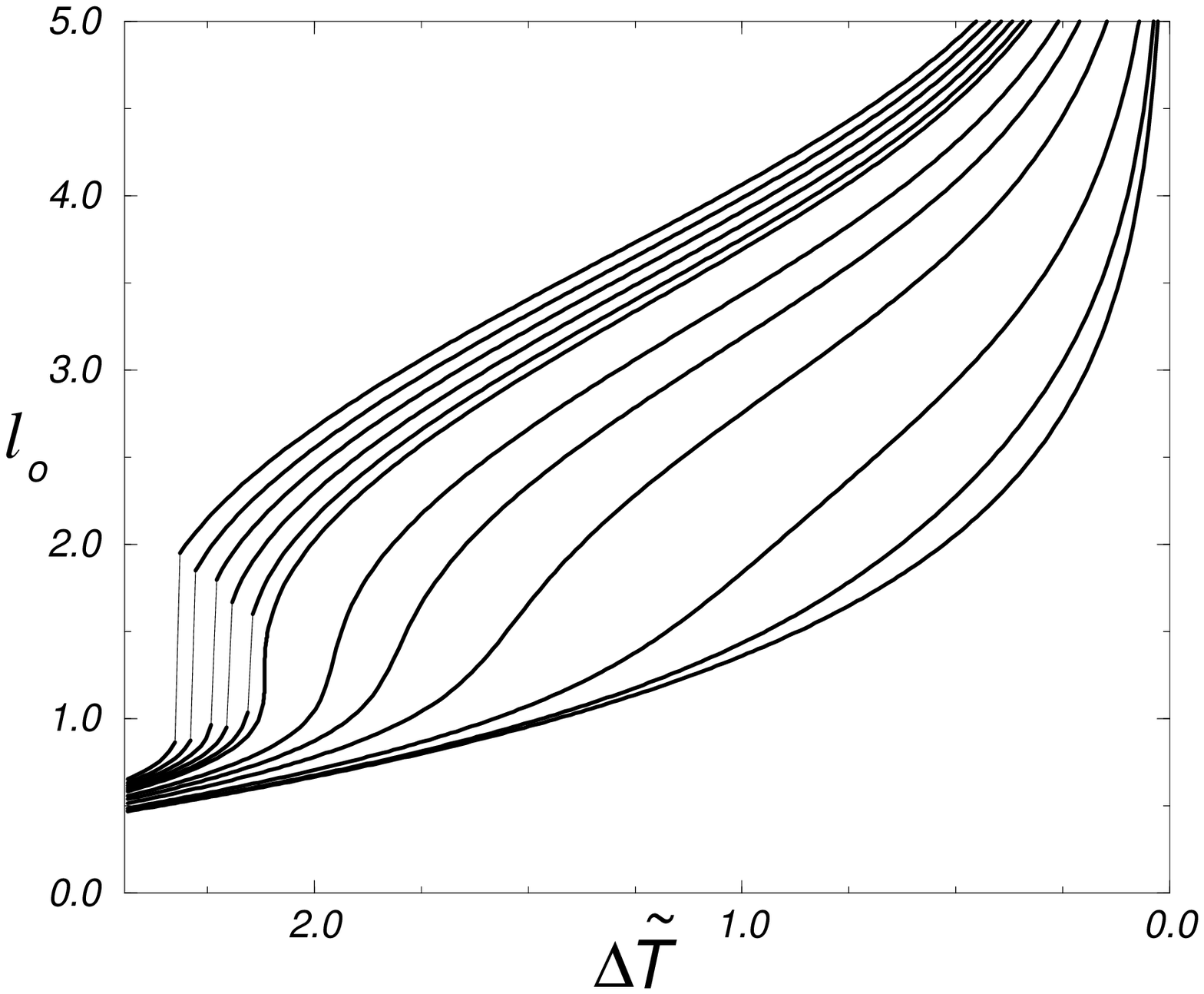,width=14cm}
\caption{Film thickness {\it vs.\ }$\Delta\widetilde{T}$
for different values of $a$ from the numerical minimization
of Eq.\ (\protect{\ref{one}}). From below,
$a/\surd{2}=$ $0.00$, $0.50$, $1.00$, $1.50$, $1.75$, $1.90$, $2.0605$,
$2.10$, $2.15$, $2.20$, $2.25$, $2.30$. All distances are measured
in units of the bulk correlation length.}
\end{figure}
\newpage

\begin{figure}[p]
\label{second}
\vspace*{3cm}
\epsfig{file=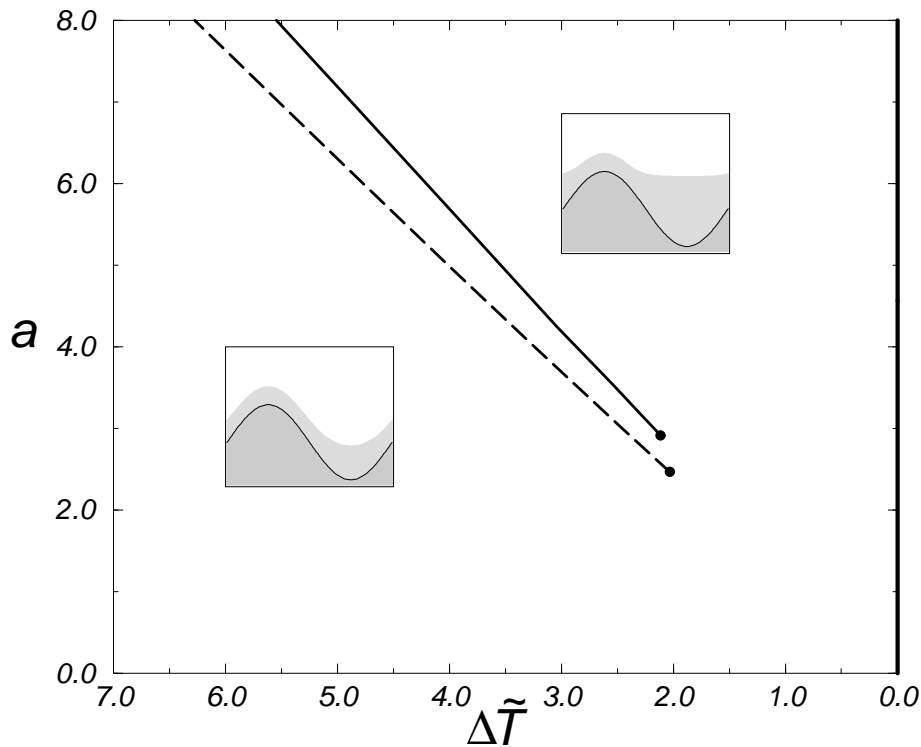,width=14cm}
\caption{Section of the surface phase diagram at bulk coexistence
showing the unbending
coexistence line which finishes at the critical point
$\Delta\widetilde{T}\approx 2.12$ and $a_{\sf c}\approx 2.914$.
The solid line represents the results of minimizing Eq.
(\protect{\ref{one}}). The dashed line is the result of
the variational approximate solution (see text).
The vertical line $\Delta\widetilde{T}\!=\!0$ represents the
second order unbinding transition. Schematic representation of the
interfacial configuration on either side of the unbending
line are also shown.}
\end{figure}
\newpage

\begin{figure}
\label{third}
\vspace*{3cm}
\epsfig{file=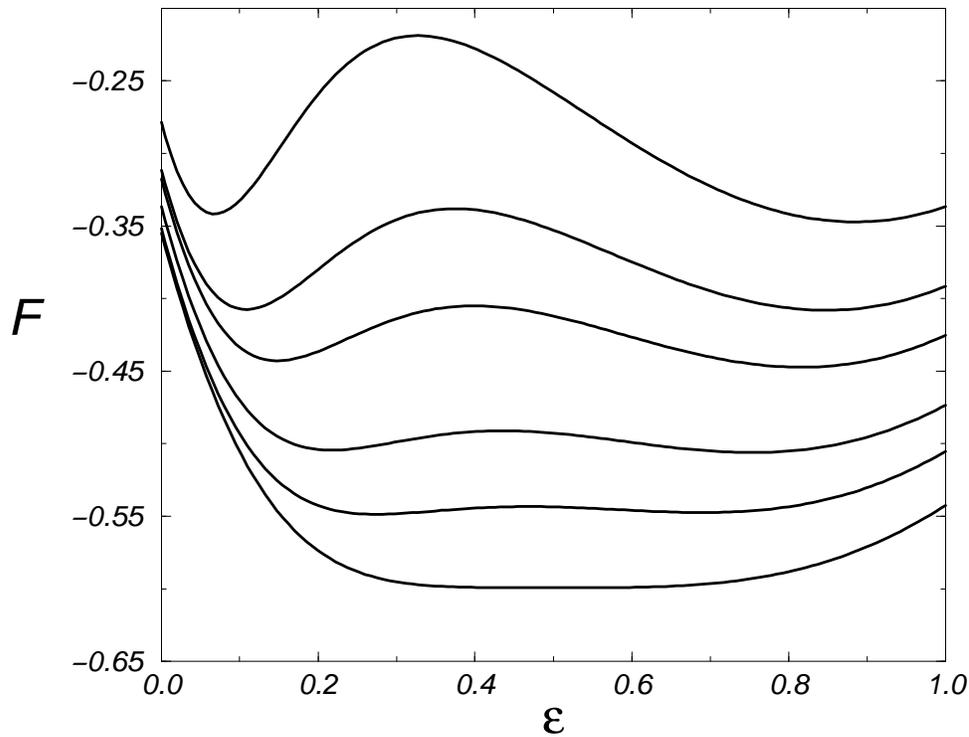,width=14cm}
\caption{Free Energy
from the expression (\protect{\ref{six}}) for differents values of $a$
at the transition temperatures (See Fig.\ 2).
From above, $a=$ $5.0$, $4.0$, $3.5$, $3.0$, $2.75$ and $2.46866$.
The last value is the critical value $a_{\sf c}$ within
the present approximation.}
\end{figure}
\newpage

\begin{figure}
\label{fourth}
\vspace*{3cm}
\epsfig{file=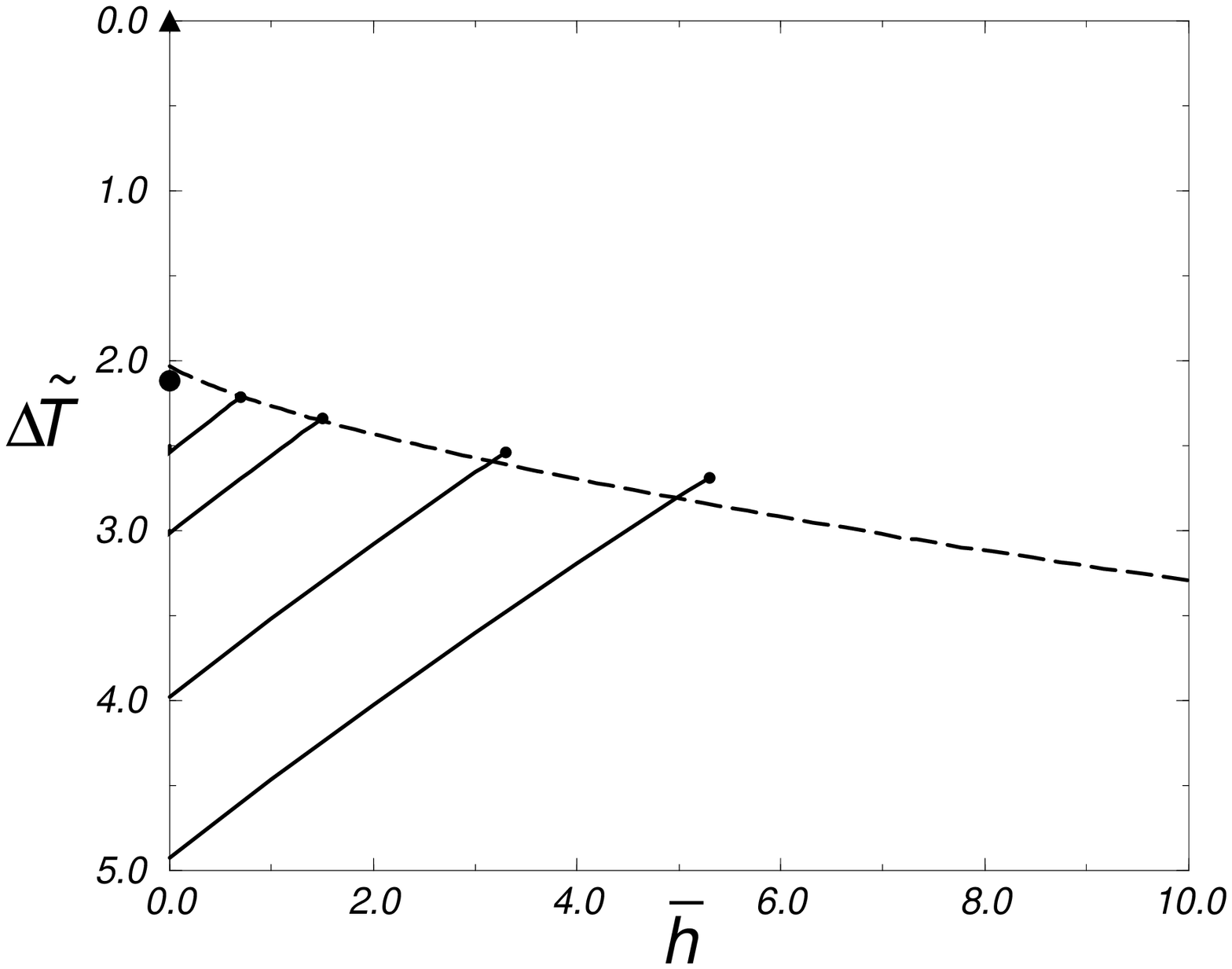,width=14cm}
\caption{Phase diagram of the unbending transition 
in a system of short ranged forces for
different values of $a/\surd{2}=$ $2.5$, $3.0$, $4.0$ and $5.0$,
from numerical minimization of Eq.\ (\protect{\ref{one}})
(continuous lines, left to right). The circle represents
the unbending critical point for $a_{\sf c}\!\approx\! 2.914$.
The loci of critical points obtained from the approximate
model is represented as a broken line. The triangle
locates the critical wetting temperature.}
\end{figure}


\begin{references}
\bibitem{Rough} For a recent review,
see for example S.\ Dietrich in Proceedings of the NATO-ASI
"New Approaches to Old and New Problems in
Liquid State Theory" (1998)
(ed., C.\ Caccamo, J.P.\ Hansen, G.\ Stell).
Other pertinent references are: 
S.\ Nechaev, Y.-C.\ Zhang, Phys.\ Rev.\ Lett.\ {\bf 74},
1815 (1995), G.\ Sartoni {\it et al.}, Europhy.\ Lett.\ {\bf 39}, 633 (1997),
T.S.\ Chow, J.\ Phys.: Cond.\ Matt.\ {\bf 10}, L445 (1998),
T.W.\ Burkhardt, J.\ Phys.\ A: Math.\ Gen.\ {\bf 31}, L549 (1998),
S.\ Curtarolo {\it et al.}, {\tt cond-mat/9808299},
P.S.\ Swain, R.\ Lipowsky, {\tt cond-mat/9809089}.
\bibitem{Dietrich} See, for example, 
S.\ Dietrich, in {\it "Phase Transitions and Critical
Phenomena"}, (C.\ Domb and J.L.\ Lebowitz, eds.), Vol.\ {\bf 12}, p.\ 1
(Academic Press, London, 1988).
\bibitem{Perturbation} (a) M.O.\ Robbins, D.\ Andelman,
J.F.\ Joanny, Phys.\ Rev.\ A {\bf 43}, 4344 (1991),
(b) M.\ Napi\'{o}rkowski, K.\ Rejmer, Phys.\ Rev.\ E
{\bf 53}, 881 (1996), (c) A.O.\ Parry, P.S.\ Swain, J.A.\ Fox, J.\ Phys.:
Condens.\ Matter {\bf 8}, L659 (1996), (d) R.R.\ Netz,
D.\ Andelman, Phys.\ Rev.\ E {\bf 55}, 687 (1997).
\bibitem{Cardy} See, for instance, {\it Finite-Size Scaling}
(ed.\ J.L.\ Cardy), North-Holland, Amsterdam (1988).
\bibitem{Equation} See, for example, R.\ Ortega, J.\ Diff.\ Eq.\
{\bf 128}, 491 (1996).
\bibitem{Bau} We anticipate that similar results to those described
here will emerge from studies of improved non-local hamiltonians.
For instance, C.\ Bauer, S.\ Dietrich (to be published).
\bibitem{Oops} Predictions, based on low-order perturbation
expansions of the free energy in terms of $a$ and $\sl q$
(see \protect{\cite{Perturbation}}(c)), that the second order
wetting transition becomes first order for sufficiently large
$a$ are not correct. Systematic inclusion of higher order terms
shows that the series expansion diverges as $\Delta T\!\rightarrow\!0$,
which is indicative of a spinodal point associated with the unbending
transition \protect{\cite{RPS}}.
\bibitem{RPS} C.\ Rasc\'{o}n, A.O.\ Parry, A.\ Sartori (in preparation).
\bibitem{Note} For systems with short-ranged forces, where
$\beta_{\sf S}\!=\!0$, the appropriate dimensionless measure of $a$ is
in units of the bulk correlation length.
\end{references}
\end{document}